\documentclass[11pt,twoside]{article}

\usepackage{asp2004}
\usepackage{epsf}
\usepackage{psfig}
\usepackage{lscape}

\begin{document}
\title{Velocity Fields at the Surface of ZZ Ceti Stars}  
\author{D. Koester and E. Kompa} 
\affil{Institut f\"ur Theoretische Physik und Astrophysik,
Universit\"at Kiel, D-24098 Kiel, Germany}  

\begin{abstract} We demonstrate that the peculiar line profiles
  observed in DA white dwarfs in the temperature range of the ZZ Ceti
  variables can be explained by the surface velocity fields associated
  with the pulsations.
\end{abstract}

\section{Introduction}  
During the spectroscopic search for rotation in white dwarfs we
noticed that all known ZZ Ceti stars showed peculiar H$\alpha$ line
profiles (Koester et al. 1998). The lines could be fitted using a
rotationally broadened profile, but the derived rotation rates were
incompatible with other data, e.g. from asteroseismology. Using the
CaII~K line for such a study, Berger et al. (2005) found rotation
velocities varying from 11-28 km/s from seven high-resolution Keck and
VLT spectra of the variable G29-38 -- a clear indication that the
broadening could not be due to rotation.

While the spectra of almost all stars could be fitted with zero
rotation, for G29-38 this was not true. The observed spectra (Fig.~1)
are much flatter than the non-rotating model. Since the rotation can
certainly not change by a factor of three within a few years, a
different mechanism has to be found.

\begin{figure}[!ht]
\plotone{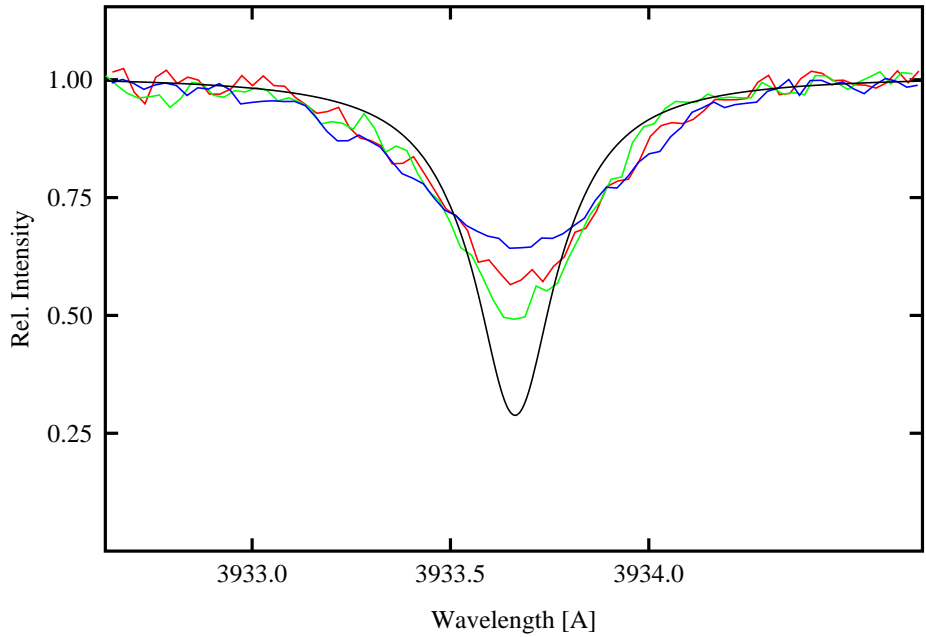}
\caption{CaII~K line in 3 high-resolution Keck spectra compared to a
  theoretical profile for a non-pulsating, non-rotating star}
\end{figure}

\section{Velocity fields}
Another obvious possibility are surface motions due to the non-radial
pulsations. Indeed, velocity shifts with the same periods as the light
variations have been observed in G29-38 (van Kerkwijk, Clemens, \& Wu 2000;
Thompson et al. 2003). In the latter study, however, these were
dismissed as a reason for the peculiar line profiles, since already
the time-resolved spectra seemed to show the same profile as the time
integrated spectrum. This argument neglects the fact, that the
velocity dispersion over the disk leads to a significant broadening
already at a single phase, compared to a non-pulsating star.

In this study we will test this conclusion by ''brute-force'' numerical
integration of time-dependent spectra over the visible disk of the
variable, taking into account temperature variations, accurate
limb-darkening, Doppler shifts from the pulsation, and the geometry of
the star in the observers frame.

The radial displacement of a surface element is described by a
spherical harmonic with mode indices $l$ and $m$. The corresponding
velocity vector is given by Dziembowski (1977). For white dwarfs the
radial component is negligible and the motions are nearly horizontal
in the stellar system.
As our test case we have chosen a $l=1$, $m=0$ mode with a period of 600 s,
and a pole-on view of the observer. After transformation of the
stellar surface velocity into the observers frame we obtain the visual
light variation, the mean velocity over the visible disk (which shifts
the line), and the velocity dispersion (which broadens the line at
each phase). Fig.~2 shows these results.

\begin{figure}[!ht]
\plottwo{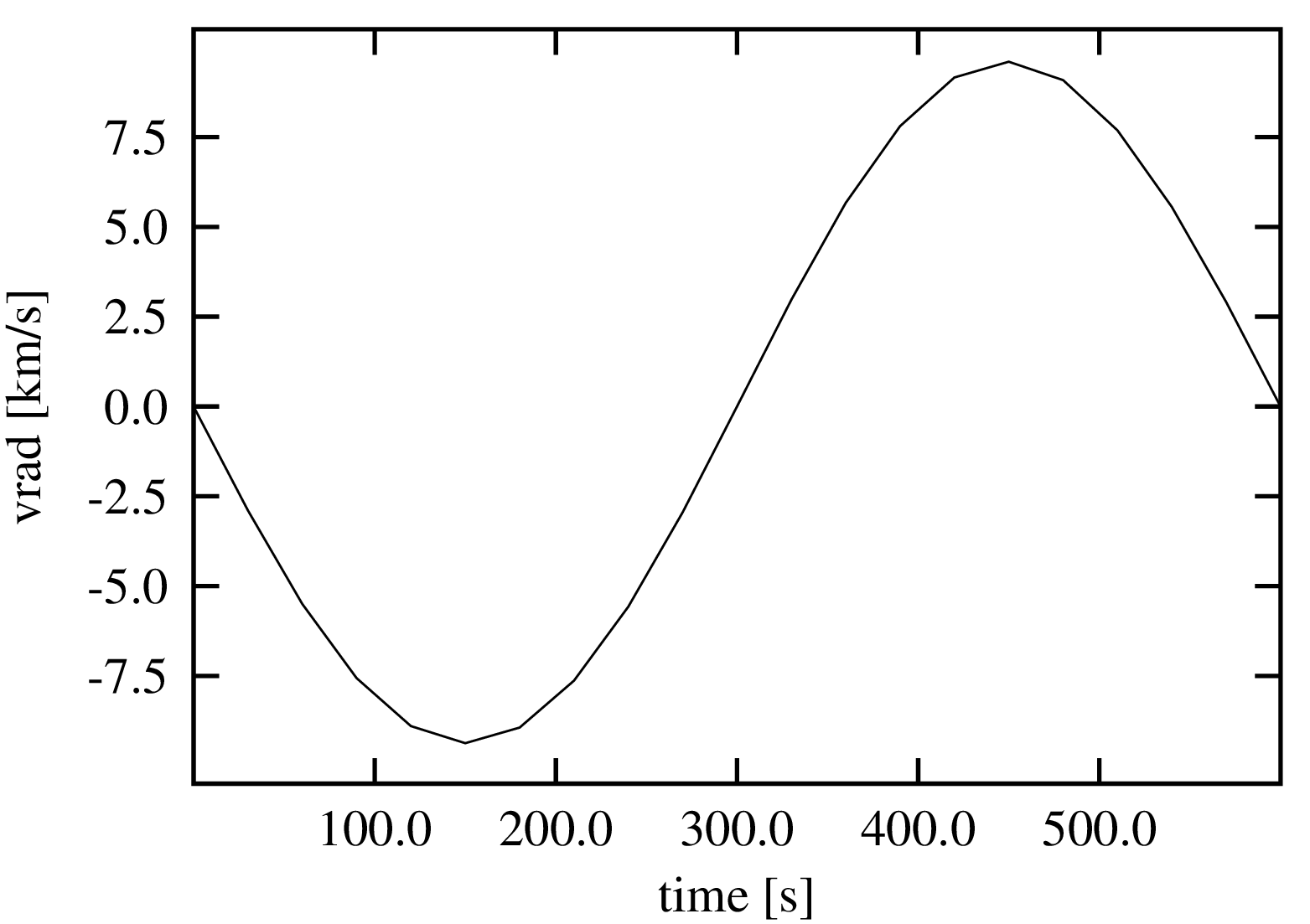}{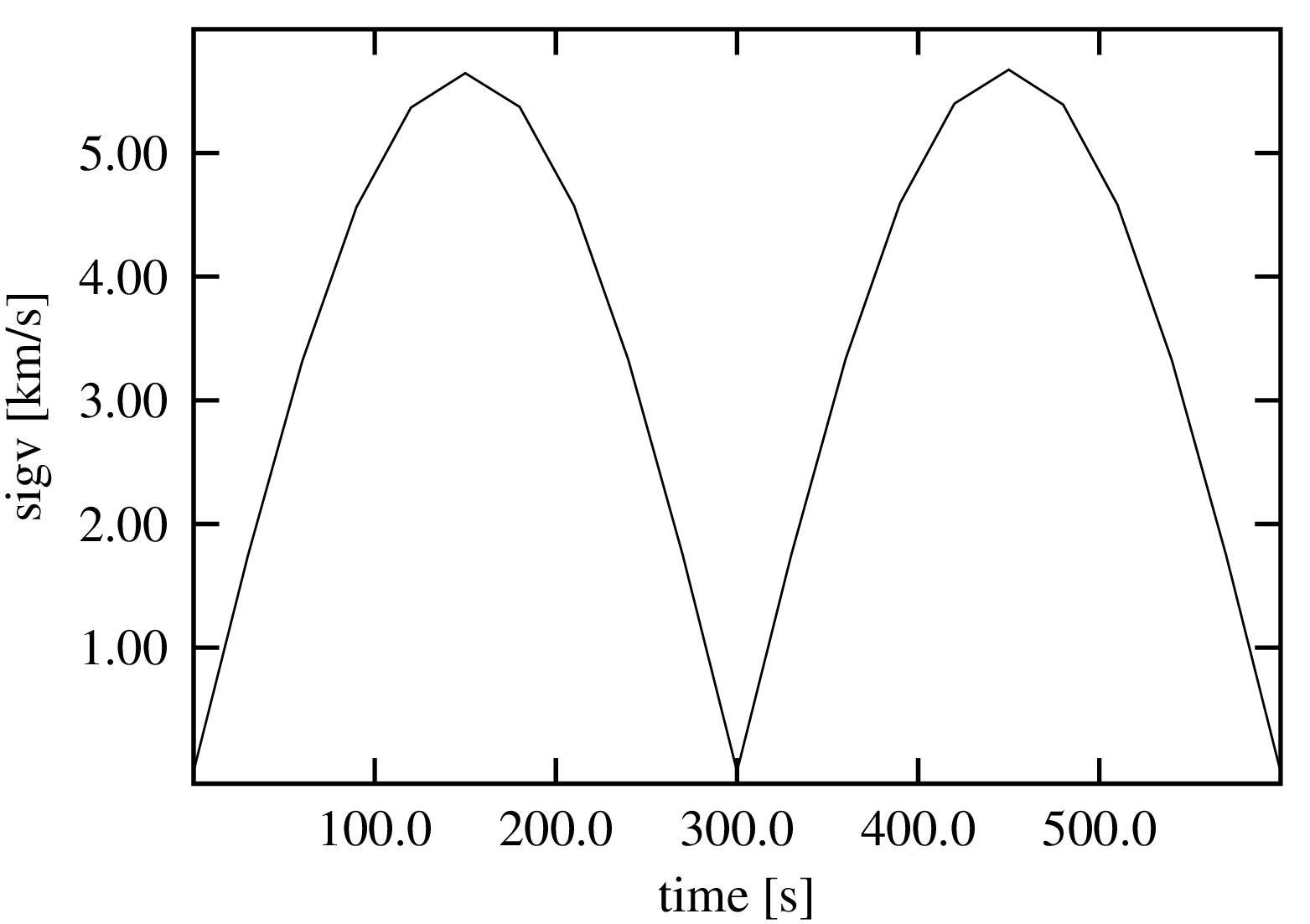}
\caption{Left: Variation of disk-averaged radial velocity over one
  period. Right: Variation of $\sigma_v$, the width of the
  velocity distribution over the visible disk.}
\end{figure}

\section{Results}
Figure~3 shows the broadening and shift of the Ca line in three
phases during the cycle. Position, width and depth of the profile
change significantly. This is the effect which would be observed with
time-resolved spectroscopy.

Figure~4 shows the comparison of three observed spectra with a profile
of a non-pulsating star and with the integrated profile (over one
period, the exposure time of Keck spectra is 1800 s) for a pulsating
star.  With the assumed values for the observable velocity amplitude
($\approx 9$ km/s) the least extreme observation can be explained. The
other spectra would need slightly larger velocities.

\begin{figure}[!ht]
\plotone{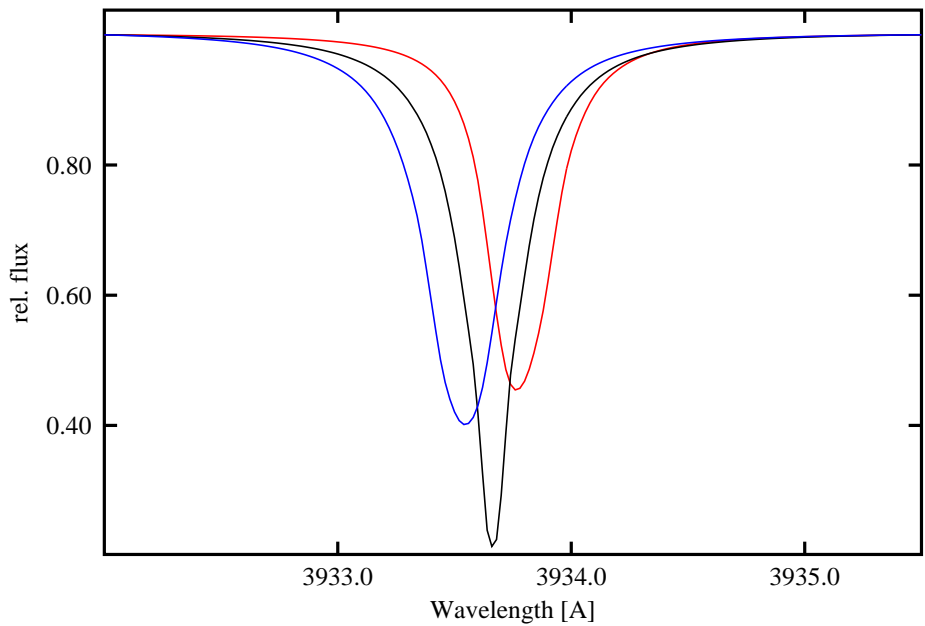}
\caption{Observable Ca line profile in three different phases of
  the pulsation cycle}
\end{figure}

\begin{figure}[!ht]
\plotone{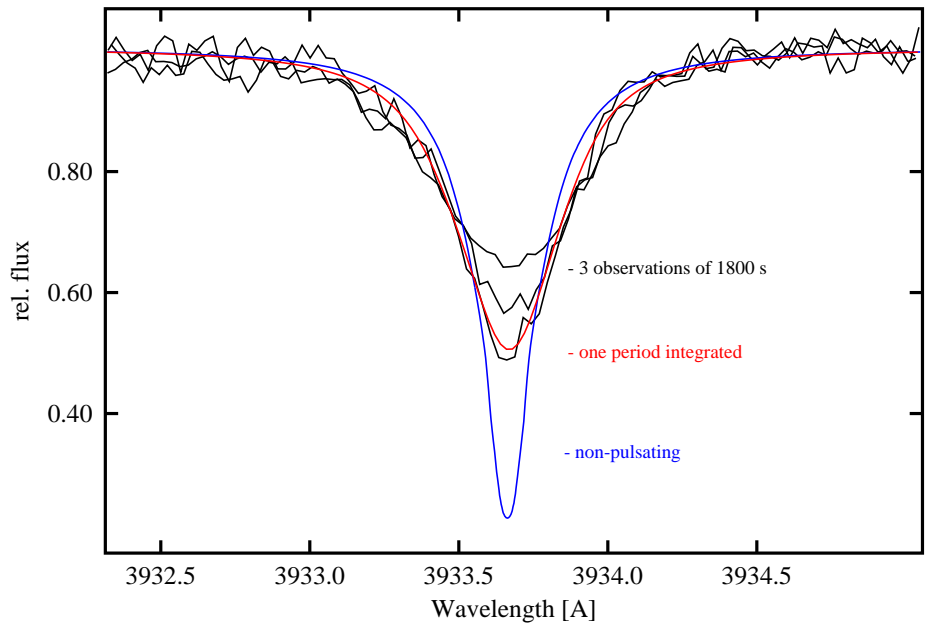}
\caption{Comparison of three observed line
  profiles (noisy curves) with the result of our simulation,
  integrated over one complete period (smooth curve, approximating
  closely the deepest observed profile. The smooth deep profile is
  calculated for a non-pulsating star.}
\end{figure}

The amplitudes we have used in this simulation for the velocity and
light variations are about a factor of 2 higher than found for a
single strong mode by Thompson et al. (2003).  However, due to several
modes present they find velocity amplitudes easily exceeding 10 km/s
in the direct measurement of the total pulsation. We are thus
convinced that these motions are at the origin of the observed
peculiar line profiles.

The study of these profiles -– unfortunately requiring very large
telescopes -– is thus another tool for the study of the variations. One
recent example is our prediction, based on the line profile, of the
variability of WD1150$-$153, which has been confirmed photometrically in
the meantime.

\acknowledgements This work in Kiel was supported by a grant from the
Deutsche Forschungsgemeinschaft (Ko739/21).

\end{document}